\RequirePackage{fix-cm}
\documentclass[smallextended]{svjour3}       
\smartqed  
\usepackage{graphicx}
\usepackage{mathptmx}      
\usepackage{color}
\usepackage{natbib}
\usepackage[hyphens]{url}
\usepackage{amsmath}
\usepackage{tabularx}
\usepackage{makecell}
\newcommand{\s}{\color{black}}
\newcommand{\black}[1]{\textcolor{black}{#1}}
\newcommand{\blue}[1]{\textcolor{black}{#1}}

\begin{document}

\title{
		On the role of data, statistics and decisions in a pandemic
}


\author{Beate Jahn$^{1, \dagger}$ \and Sarah Friedrich$^{2, \dagger}$ \and Joachim Behnke$^3$ \and Joachim Engel$^4$ \and Ursula Garczarek$^5$ 
	 \and Ralf M\"unnich$^6$ \and Markus Pauly$^7$ \and Adalbert Wilhelm$^{8}$ \and Olaf Wolkenhauer$^{9}$ \and Markus Zwick$^{10}$ \and Uwe Siebert$^{1, 11, 12}$ \and  Tim Friede$^2$\footnote{Corresponding author: 
  \email{tim.friede@med.uni-goettingen.de}}
\thanks{$^\dagger$ Shared first authorship}
}


\institute{$^1$ Institute of Public Health, Medical Decision Making and Health Technology Assessment, Department of Public Health, Health Services Research and Health Technology Assessment, UMIT – University for Health Sciences, Medical Informatics and Technology, Austria \\
      \and
      $^2$ Department of Medical Statistics, University Medical Center G\"ottingen, Germany\\
        \and $^3$ Zeppelin University Friedrichshafen, Germany \\
     \and $^4$ Pädagogische Hochschule Ludwigsburg, Germany\\
    \and $^5$    Cytel Inc, 675, Massachusetts Avenue, Cambridge, MA 02139 USA \\
     \and $^6$ Economic and Social Statistics, Trier University, Germany \\
    \and $^7$   Department of Statistics, TU Dortmund University, Germany\\
   \and $^{8}$    Psychology and Methods, Jacobs University Bremen, Germany \\
   \and $^{9}$  Department of Systems Biology \& Bioinformatics, University of Rostock and Leibniz-Institute for Food Systems Biology, Technical University of Munich, Germany\\
      \and $^{10}$ Goethe University Frankfurt, Germany\\
  \and $^{11}$ Institute for Technology Assessment and Department of Radiology; Massachusetts General Hospital; Harvard Medical School, Boston, MA, USA\\ 
   \and $^{12}$ Center for Health Decision Science and Departments of Epidemiology and Health Policy \& Management, Harvard T.H. Chan School of Public Health, Boston, MA, USA
  }

\date{Received: date / Accepted: date}

\maketitle

\begin{abstract}
{\color{black}
A  pandemic  poses  particular  challenges  to  decision-making because of the need to continuously adapt decisions to rapidly changing evidence and available data. For example, which countermeasures are appropriate at a particular stage of the pandemic? How can the severity of the pandemic be measured? What is the effect of vaccination in the population and which groups should be vaccinated first? The process of decision-making starts with data collection and modeling and continues to the dissemination of results and the subsequent decisions taken. The goal of this paper is to give an overview of this process and to provide recommendations for the different steps from a statistical perspective. In particular, we discuss a range of modeling techniques including mathematical, statistical and decision-analytic models along with their applications in the COVID-19 context. With this overview, we aim to foster the understanding of the goals of these modeling approaches and the specific data requirements that are essential for the interpretation of results and for successful interdisciplinary collaborations. A special focus is on the role played by data in these different models, and we incorporate into the discussion the importance of statistical literacy, and of effective dissemination and communication of findings.
}

\keywords{COVID-19 \and SARS-CoV-2 \and Health-decision framework \and Decision-analytic modeling}
\end{abstract}


\section{Introduction}
\label{intro}

In December 2019, the first cases of coronavirus disease 2019 (COVID-19) were reported in Wuhan, China {\color{black} \citep{zhou2020pneumonia,wu2020new}} and the outbreak of severe acute respiratory syndrome coronavirus 2 (SARS-CoV-2) was declared a pandemic in March 2020 by the World Health Organization. In order to control the spread of the virus and limit the negative consequences of the pandemic, important decisions had and still have to be \black{made}. These concern the spread of the disease, its impact on health, the utilization of health care resources or potential effects of counter measures \black{and vaccination strategies}, to name some examples.
\black{Statistical modeling plays an important role in different fields of COVID-19 research. This starts with the \blue{collection of adequate data and the preprocessing of this data}, a complex sequence of steps, where input is required from the data users, taking into account their questions and information needs. After this preprocessing, examples of statistical models range from characterizing the disease \citep{kuchenhoff2021analysis,roy2021spatial,luo2021time}, investigating comorbidities \citep{gross2021covid,hadzibegovic2021heartfailure,Evangelou107},
evaluating new treatments and vaccines with respect to efficacy and safety \citep{horby2020lopinavir,recovery2020effect,nejm2021vaccine,flaxman2020estimating} as well as planning corresponding trials \citep{MUTZE2020106154,stallard2021adaptive,beyersmann2021covid}, assessing the spread of the disease in potential scenarios – such as comparing lockdown or vaccination strategies \citep{nussbaumer2020quarantine,Pelt2021,Jahn2021} – and evaluating the impact of the pandemic on clinical trials \citep{kunz2021,anker2020}. One important aspect in the special situation of a pandemic \blue{with a novel pathogen} is the incorporation of sequential inference, that is, continuously updating the research as new data becomes available.} 

In the course of the pandemic, the availability and quality of data, the varying interpretations of modeling results, as well as apparently contradicting statements by scientists, have caused confusion and fostered intense debate. The role, use and misuse of modeling for infectious disease policy making has been critically discussed \citep{james2021use,holmdahl2020wrong}. Furthermore, the CODAG reports \citep{codag-reports} clarify why models can lead to conflicting conclusions and discuss the purposes of modeling and the validity of the results.
For instance, policies to contain the pandemic were – {\color{black} in the beginning} – mainly guided by 7-day incidence. 
Measures such as curfews, limited numbers of guests at events and restricted opening hours of stores were driven by this figure. However, considering the 7-day incidence alone does not provide a meaningful view of the overall picture as discussed by \cite{codag}. As mentioned in the series ``Unstatistik'' \citep{Unstatistik202010} a value of 50 cases per 100,000 inhabitants in October 2020 in Germany had an entirely different meaning than six months earlier due to changes in testing strategies, and improved treatments among other factors. Concerning the expected number of intensive care patients and deaths, a value of approximately 50 in October 2020 is likely to correspond to a value of 15 to 20 in April 2020, possibly even less \citep{Unstatistik202010}.
\black{Recently, the hospitalization and ICU incidences have been considered as additional measures. While this provides a more reliable picture of the severity of the situation and is less affected by differing testing strategies, it is not without shortcomings. \blue{For example,} under-reporting and time-lags lead to large differences between reported and actual numbers. \blue{Moreover, as the severity of COVID infections dropped with the Omicron variant and prevalence increased, a new discussion of hospitalization ``with'' or ``because of'' COVID-19 emerged. These examples highlight} the need to use statistical methods such as nowcasting \citep{gunther2021nowcasting,schneble2021nowcasting,salas2021improving,altmejd2020nowcasting} for more precise estimations.}

As known from the field of evidence-based medicine and health data and decision science, decisions should be underpinned by the best available evidence. For evidence-based decision making, three components are important: a) data, b) statistical, mathematical and decision-analytic models (which reduce the amount and the complexity of the data to meaningful indices, visualizations, and/or predictions), and c) a set of available decisions, interventions or strategies with their consequences described through a utility or loss function (decision-making framework) and the related tradeoffs. 
\color{black} \black{General international guidance on these assessments and decision analysis is implemented in a  country-specific manner, mainly by Health Technology Assessment (HTA) organisations \citep{Drummond2008, Gandjour2020}. COVID-19 examples include the evaluation of vaccination strategies \citep{Kohli2021, Debrabant2021, reddy} or treatment \citep{Sheinson2021}}.
Scientists of the German Network for Evidence-Based Medicine raised the question on ``COVID-19: Where is the evidence?'' \citep{EBM2020} which motivated a discussion about the need of randomized controlled trials to investigate the effectiveness of preventive measures, feasibility of such studies and longitudinal, representative data generation.


{\black{In the pandemic, a multitude of models have been used but the systematic comparison across different classes of models is lacking. The goal of this paper is to provide an overview of the  process from  data collection (primary and secondary) and modeling, up to communication and decision making and to provide recommendations related to these areas.
We discuss a range of modeling techniques including mathematical, statistical and decision-analytic models along with their application in the COVID-19 context. With this overview, we aim to foster the understanding of the goals of these modeling approaches and the specific data requirements that are essential for the interpretation of results and a successful interdisciplinary collaboration.
Model types less known to statisticians, such as decision-analytic models, still require statistical thinking. In particular, 
functional relationships and input parameters for these models are often provided by statisticians and epidemiologists. Our target audience, therefore, is broad. It includes data scientists – such as
statisticians – mathematicians, physicists, epidemiologists, economists, social and computer scientists, and decision scientists.}\\

The paper is organized as follows. In Section~\ref{sec:overview} we give a short overview of modeling purposes and approaches with a special focus on differences between disciplines. In Section~\ref{sec:data}, we discuss requirements of data quality and why this is fundamental for the entire process. We then move on to modeling, with Section~\ref{sec:models1} dealing with the different purposes of modeling. In Section~\ref{sec:decision}, we explain how decisions can be informed based on these models. We discuss aspects of the reporting and communication of results in Section~\ref{sec:reporting} {\color{black}, provide recommendations in Section~\ref{sec:recom} and a discussion in Section~\ref{sec:discussion}}.

}

\section{\black{Overview of modelling approaches and their purposes}}\label{sec:overview}

\begin{figure*}
	\includegraphics[width = \textwidth]{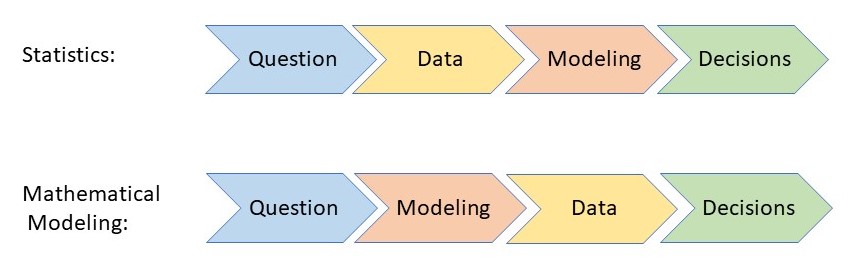}
	\caption{For evidence-based decision making, a complex process is often necessary. In both statistical and mathematical/decision-analytic modeling, the path starts with a research question and ends in guidance for decision making. In statistics (upper path) using data as the basis for modeling is common. In mathematical and decision-analytic modeling (lower path), models are based on subject matter knowledge and validated or calibrated using data sets. {\color{black} For the purpose of illustration, the process is depicted as a sequence of steps here. In reality, however, these are more likely to be cyclic, iterating steps such as data capturing and modeling and informing new questions based on previous results.}} 
	\label{fig:process}
\end{figure*}

As a statistician, the process of gaining knowledge starts with a research question and continues with the acquisition of data, which then enters into the statistical model – see Figure 1 in \cite{DAGStat2020} for an illustration. Data acquisition here might refer either to the design of an adequate experiment or observation, or to the use of so-called secondary data, which has been collected for a different purpose. Statistical principles of design are relevant, even when using secondary data \citep{rubin2008}. In Bayesian statistics, other information can be incorporated as \emph{a priori information} \citep[e.g.][]{ohagan2004bayesian}. This might stem from previous studies or might be based on expert opinions. The prior information combined with the data (likelihood) then results in a posterior distribution. 
In modeling contexts outside statistics, data and/or information is used in different ways. Simulation models use prior information (again based either on data or on other sources such as expert opinion or beliefs) to determine the parameters of interest and are usually validated on a data set. The formal representation of the mathematical or decision-analytic model makes assumptions about the system that generates the data \black{\citep{Roberts2012}}, and the (mis)match between data and model then provides insights that can be used as basis for decisions. In contrast to statistical models, the order of data and modeling is thus reversed. An illustration is depicted in Figure \ref{fig:process}. 
For the purpose of illustration, Figure \ref{fig:process} depicts the process as a sequence of steps. As the pandemic progresses, however, some steps such as data capturing and modeling might be iterated.

{\color{black} From a mathematical perspective, a statistic $f$ is a quantity or function defined on the sampling space. 
Note that the term statistic is used both for the function $f$ as well as for the value $f({\bf x})$ of the function on a given data set ${\bf x}$ \citep{degroot,mcgraw}.
Choices of $f$ include very simple preprocessing steps, e.g., taking the mean where $f(x_1, \dots, x_n)= \frac{1}{n}\sum_{i=1}^n x_i$, as well as estimates obtained from complex statistical models, such as hierarchical Bayesian models used for nowcasting \citep{gunther2021nowcasting} or sophisticated regression models for prediction \citep{iwendi2020covid}. 
Thus, the input to a statistical model might either be `raw data' or might have undergone some previous steps, like scaling or transformation applied to variables before entering them in a regression model. These previous steps are often referred to as preprocessing (especially in the context of machine learning) and examples include descriptive statistics and exploratory data analysis, see also \cite{DAGStat2020} and the references cited therein. It should be noted, however, that data can often not be analyzed directly, but is the product of a complex sequence of processing steps. In particular, \cite{desrosieres2010politics} separates three aspects of statistics, namely ``(1) that of quantification
properly speaking, the making of numbers, (2) that of the uses of numbers as variables, and finally, (3) the prospective inscription of variables in more complex constructions, models''.
Throughout this paper, we refer to the model input parameters as `data' irrespective of whether some preprocessing took place or not. Official statistics, for instance, usually refer to crude observations when talking about `data', while the information obtained by (pre-)processing this data is called statistics. 
For this paper, however, we adapt a slightly different view on the same aspect and do not distinguish between crude and preprocessed data. In this sense, the measures of quality and trustworthiness discussed in Section~\ref{sec:data} similarly extend to preprocessing, descriptive statistics and exploratory data analysis \citep{DAGStat2020}.}


As mentioned above, modeling is an integral part of evidence-based decision making. Here, we distinguish three purposes of modeling, which are summarized in Table~\ref{tab:1}.
The first category contains models which aim to explain patterns and trends in the data. The second category aims to predict the present (so-called now-casting) or the future (forecasting). Finally, decision-analytic models aim to inform decision makers by simulating the consequences of interventions and their related tradeoffs (e.g. benefit-harm tradeoff, cost-effectiveness tradeoff).  

	\begin{table}
		\caption{Overview of modeling purposes and approaches. For each different group of models (columns) we describe goals, approaches and challenges.}
		\label{tab:1}       
		\begin{tabularx}{\linewidth}{ | c | X | X | X | }
			\hline \noalign{\smallskip}
		{\bf Purpose:}	& {\bf Explanation} & {\bf Prediction} & {\bf Decision Analysis}  \\ \hline \hline
			Goal &		Explain patterns, trends and interactions. 
			& Predict present (e.g.~$R$ value)
			or future (e.g.~ICU bed capacity)
			& Evaluate predefined alternative interventions, actions or technologies (e.g.~vaccination strategies)
				\\ \hline
			Focus & Dynamic	patterns & Absolute	numbers & Ranking of size and direction	of effect, \black{quantifying tradeoffs} \\ \hline
	\makecell{Statistical	approaches \\ (Examples)}&
			Factor analysis, cluster analysis, regression, contingency analyses, {\color{black}propensity score methods, g-computation, marginal structural models}
			&
			Time series analysis, repeated measure analysis, {\color{black} statistical and} machine learning.
			& Statistical decision theory\\ \hline
		\makecell{Simulation models\\ (Examples)} & 
			ABM$^*$, SD$^*$, differential
			equations, {\color{black} MSM$^*$} & 
			Differential equations, ABM$^*$, {\color{black} MSM$^*$} & 	
			 Differential equations, ABM$^*$, MSM$^*$, SD$^*$, DES$^*$,
			State-Transition Models,
			Decision Trees\\
			\hline
			
			\makecell{Importance of \\ correctly assessing \\
				quantitative \\ uncertainty} &
			Minor role (+) &
			Major role (+++) &
			Intermediate (++) \\ \hline
			\makecell{Importance of \\ correctly assessing \\ qualitative \\ uncertainty} &
			Intermediate (++) &
			Minor role (+) &
			Major role (+++)\\ \hline
			\makecell{Weaknesses/ \\
				Challenges} &
			High risk of overinterpretation w.r.t.~causality
			
			& 
			Highly sensitive to context
						
			& Highly dependent on comprehensive choice of
				options and outcomes \\
			
			\noalign{\smallskip}\hline
		\end{tabularx}
		{\raggedright ABM Agent-based model, DES discrete event simulation, SD system dynamics, ICU intensive-care unit, MSM Microsimulation Modeling \par}
	\end{table}
	 
In medical decision making and health economics, this leads to a formal decision framework, which relates to statistical decision theory \citep{siebert2003should,siebert2005using}.
In this framework, \citep{parmigiani2009decision} the decision maker has to choose among a set of different actions. The consequences of these actions depend on an unknown ``state of the world''. The basis for decision making depends on the quantitative assessment of these uncertain consequences. To this end, a loss or utility function must be defined which allows the quantification of benefits, risk, cost or other consequences of different actions. Minimizing the loss function (or maximizing the utility function) then leads to optimal decisions.

\section{Data availability and quality}
\label{sec:data}

An essential basis for research and for evidence-based policy is high quality data. The mere presence of data is not enough, as the process of data definition, collection, and processing determines the quality of the data in reflecting the phenomena on which to provide evidence. Poor definition of data {\s concepts and variables as well as bad choices in their collection and processing can lead to misleading data, that is }data with severe bias, or to {\s an unacceptably large} remaining level of uncertainty about the phenomena of interest, so that any results generated with that data form an inadequate basis for decision making. 
Below, we describe which quality characteristics have to be considered when planning a data collection or when assessing the quality of already existing data for a task at hand. The underlying concepts are general and well-known. Despite this, they are regretfully often neglected, and thus we summarize them as eight characteristics. in the context of policy making.  

\begin{enumerate}
	\item \textbf{Suitability for a target:} Data in itself is neither good nor bad, but only more or less suitable for achieving a certain goal. In order to assess data, it is first necessary to understand, agree on, and describe the goal that the data are supposed to support. 
	\item \textbf{Relevance:} Data must provide relevant information to achieve the goal. To do this, the data must measure the characteristics needed (e.g.~how to measure population immunity?) on the right individuals (e.g.~representative sample for generalization, or high-resolution data for local action?).
	\item \textbf{Transparency:} The data collection process must be transparent in terms of origin, time of data collection and nature of the data. Transparency is a requirement for peer-review processes to ensure correctness of results, and for an adequate modeling of uncertainties. 
	\item \textbf{Quality standards:} Data are well suited for policy making requiring general overviews and spatio-temporal trends if local data collection follows a clear and uniform definition of what is recorded and how it has been recorded. Standardization includes, for example, the harmonization of data processes, adequate training of the persons involved in the collection, and monitoring of the processes.
	\item \textbf{Trustworthiness:} To place trust in the data, these must be collected and processed independently, impartially and objectively. In particular, conflicts of interest should be avoided in order not to jeopardize their credibility.
	\item \textbf{Sources of error:} Most data contain errors, such as measurement errors, input errors, transmission errors or errors that occur due to non-response. With a good description of data collection and data processing (see `Transparency' above), possible sources of error can be assessed and incorporated into the modeling for the quantification of uncertainty and the interpretation of results.
	\item \textbf{Timeliness and accuracy:} Ideally, data used for policy-making should meet all quality criteria. However, information derived from data must additionally be up-to-date and some decisions (e.g.~contact restrictions) cannot be postponed to wait until standardized processes have been defined and implemented and optimal data has been collected. The greater uncertainty in the data associated with this must be met with transparency and with great care in its interpretation. 
	\item \textbf{Access to data for science:} In order to achieve the overall goal of evidence-based policy making, it is important to make good data available as a resource to a wide scientific public. This allows for the data to be analyzed in different contexts and with different methods, and enables the data to be interpreted from the perspective of different social groups and scientific disciplines.
	
\end{enumerate}

\black{These eight aspects are included in the European Statistics Code of Practice \citep{european}. This Code of Practice, however, goes beyond the above mentioned points by covering further aspects of statistical processes and statistical outputs as well as an
additional section on the institutional environment.
The aim is to provide a common quality framework of the European Statistical System.}


The items \black{presented above} mainly refer to a primary data generating process, that is, when the data are directly generated to provide information on a pre-defined target. Especially in the context of COVID-19, information from \emph{available sources} often has to be considered, where the data generation does not necessarily coincide with the aim of the study. One prominent example is the number of infections, which are gathered by the local health authorities, but are used for comparing regional incidences which are the basis of several policy decisions. Particular attention in this case has to be paid to selection bias. One way to assess (and thus address) selection bias would come from accompanying information on asymptomatically infected persons gained through representative studies. Other information to mitigate selection bias comes from the number of tests and the reasons for testing, but these are not appropriately reported in Germany. Both problems yield biased regional incidences. Hence, modeling based on these data may cause misleading results and has to be considered carefully. Additionally, the data generating process may be subject to informative sampling \citep{pfeffermann2009inference}. 

The above aspects always have to be seen in light of the research question. Incidences and infection patterns need highly different data. Available data are often inappropriate, or must be accompanied by additional data sources. Due to the highly volatile character of COVID-19 infections, data gathering – especially via additional samples – must be very carefully planned to foster the necessary quality to provide the foundation for policy actions \citep{rendtel2021}.

Representativity implies drawing adequate conclusions from the sample on the population or parameters of the population. To achieve this, known inclusion probabilities on a complete list of elements must be given in order to allow statistical inference. Nowadays, the term representativity is generalized to cover regional smaller granularity, as well as in accordance with the time scale. Further details, especially for subgroup representativity, can be drawn from \citet{gabler2013reprasentativitat}. In household or business surveys, the term representativity has to be seen in the context of non-response and its compensation  \citep{schnell2019survey}. In practice, the term representativity is often recognized as a sufficiently high quality sample. This is entirely misleading. Indeed, statistical properties, and especially accuracy, have to be additionally considered \citep{munnich2020qualitat}. Finally, it has to be pointed out that these aspects have to be separately considered for each variable or target of interest.

\section{From data to insights -- the purposes of modeling}
\label{sec:models1}

Data and decisions are often linked using statistical models or simulations. {\color{black} As noted in Section \ref{sec:overview}, we refer to data as the input to statistical models irrespective of any preprocessing steps. As such, data preparation, descriptive statistics and preprocessing are not the focus of this paper and are thus not discussed in detail. Nonetheless, they are an essential step in any statistical analysis and especially in a situation such as the COVID-19 pandemic, where data are, in particular in the early stages of a pandemic, sparse and often disorganized. Exploratory data analysis is an important step to check data quality and discover potential anomalies in the data. An important aspect, however, is to keep in mind that data is a product of a complex sequence of steps. Here, transparency concerning the origin of the data and the whole process of data preparation is important and should be included as meta-data.}
As {\color{black} outlined} in the Section \ref{sec:overview}, modeling can serve three purposes. Each of them can be approached from either a statistical or a mathematical modeling perspective. {\color{black} In these models,} data can play different roles: while statistical models use data as the basis for the model itself, simulations are based on parameters according to prior information and predictions based on the simulation can be checked against real data to assess the precision and validity of the constructed model.

An important aspect is handling and communicating uncertainty. In statistical models, different types of uncertainty occur: sampling variation, model uncertainty, incomplete data, applicability of information and confounding are common examples \citep[e.g.][]{altman2014uncertainty,abadie2020sampling,chatfield1995model}.
In mathematical and decision-analytic models, there are usually alternative approaches to determine the values for key parameters used in simulations. For decision-making purposes, therefore, it is important to compare different methods for determining the indicators. 
Consequently, in most cases, not a single number but an interval or distribution has to be considered. 
{\color{black} In the following, we will consider the three purposes of modeling in more detail. For each of them, we provide some mathematical background, explain the difference between statistical models and mathematical or decision-analytic models and give some examples of how these approaches were applied in the COVID-19 pandemic.}

\subsection{Modeling for explanation} \label{sec:explain}

The main goal of these models is to explain patterns, trends or interactions. 
Statistical models for this purpose include, for example, regression models \citep{fahrmeir2007regression} as well as factor, cluster or contingency analyses \citep[e.g.][]{fabrigar2011exploratory,duran2013cluster}. 
In this context, associations are often misinterpreted as causal relationships. The discovery of correlations and associations, however, cannot be equated to establishing causal claims. 
In statistics and clinical epidemiology, for example, the Bradford Hill criteria \citep{hill1965} can be used to define a causal effect.

From a statistical perspective, there are two possibilities to tackle this issue. The gold standard is to design a randomized experiment, which enables causal conclusions. 
In the context of the SARS-CoV-2 pandemic, randomized controlled trials were used for assessment of COVID-19 treatments including the RECOVERY platform trial leading to publications such as \cite{horby2020lopinavir,abani2021convalescent,recovery2020effect}. In the development of vaccines, too, randomized controlled trials {\color{black} (RCT)} played a vital role \citep[e.g.][]{baden2021nejm,nejm2021vaccine}.
{\color{black} However, randomized experiments are not always feasible due to ethical considerations, cost constraints, and other reasons. Moreover, RCTs have been criticised for a number of problems including their lack of external validity \citep{rothwell2005external} or the Hawthorne effect \citep{mayo2004human}.}

Where randomized experiments are not possible and observational data is used instead, causal conclusions are harder to draw. In order to get valid estimates in this situation, a common approach is the counterfactual framework by \cite{Rubin1974}. For simplicity, assume that we are interested in the effect of a binary ``treatment'' $A \in \{0, 1\}$ (this could be an ``{\color{black} immediate} lockdown'' vs. ``no {\color{black} immediate} lockdown'', for example) on some outcome $Y$ (e.g.~number of infections with COVID-19). Then we denote $Y^{a=1}$ as the outcome that would have been observed under treatment $a=1$ and $Y^{a=0}$ the outcome that would have been observed under no treatment ($a=0$). A causal effect of $A$ on $Y$ is now present, if $Y^{a=1} \neq Y^{a=0}$ for an individual. In practice, however, only one outcome can be observed for each individual. Thus, it is only possible to estimate an average causal effect, i.e.~$E(Y^{a=1}) – E(Y^{a=0})$ \citep{Hernan2020causal}. Different possibilities for estimating a causal effect have been proposed, for example, propensity score methods \citep{Cochran1973}, the parametric g-formula \citep{robins2004}, marginal structural models \citep{robins2000marginal}, structural nested models \citep{robins1998structural} and graphical models \citep{didelez2007graphical}. Recent works have shown that these methods have difficulties when it comes to small sample studies as in the context of COVID-19 \citep{friedrich2020causal}.
{\color{black} Note that the methods explained here can also be applied to more complex situations such as non-binary treatments. In the pandemic, for example, it might be relevant to compare different time points for starting the lockdown, i.e., to include a time dimension in the considerations above. Furthermore, the methods can also be extended to more complicated outcome variables, e.g.~time-to-event data.}

Mathematical models and simulations can also be used to understand and explain dynamic patterns. Examples are simulation studies for public health interventions such as lockdown and exit strategies, where general consequences of different measures can be compared. Seemingly simple simulation models have played an important role in communicating the dynamics during a pandemic.
In these models, assumptions about the system that generates the data and (causal) relationships are made. The (mis)match between data and model then provides insights that can be used as basis for decisions. However, this procedure does not establish causal relationships in the statistical sense described above.
	
The main challenge in modeling for explanation is good communication, irrespective of whether the model is based on statistical or mathematical approaches.
{\color{black} Therefore, we consider standards for good communication in detail in Section \ref{sec:reporting}.}
Anticipating the human bias for interpreting results causally, clear statements need to be made to which extent (from ``not at all'' to ``plausible'') specific detected associations allow some causal interpretation and why. The two extreme interpretations – on the one hand, the simple disclaimer that ``correlation is not causation'', on the other, blanket and unqualified causal interpretations – do a disservice to the complexity of the problem as outlined by the methods above. 









\subsection{Modeling for prediction} \label{sec:prediction}

In statistical prediction models, the modeler can choose from large toolboxes in (spatio)-time-series analysis as well as statistics and machine learning (ML). Examples cover simple but interpretable ARIMA models \citep{benvenuto2020application,roy2021spatial}, support vector machines \citep{rustam2020covid}, joint hierarchical Bayes approaches \citep{flaxman2020estimating} or state-of-the art ML methods, such as long short-term memory (LSTM) or extreme gradient boosting (XGBoost) \citep{luo2021time}. A comprehensive overview is also given by \cite{kristjanpoller2021causal}.

For predictions based on such models, one can distinguish different aims: now-casting and forecasting. 
For \emph{now-casting}, information up to the current date and state are used to estimate or predict key figures, like the $R$ value, for example, which estimates during a pandemic how many people an infected person infects on average. 
In \emph{forecasting}, spatio-temporal predictions or simulations are used to look ahead in time, as in a weather forecast, or to estimate the required number of ICU beds.
\black{An important aspect in this situation is that the behavior of people influences the process that is being modeled. To be concrete, policy decisions are based on the predictions of a statistical or mathematical model and by introducing certain counter-measures, the original predictions of the model never come true. Thus, these models are not prediction models in a classical sense but more like projections, i.e.~scenarios of what would happen if no intervention was taken. A thorough discussion of this topic can be found in \cite{hellewell}.
}
In {\color{black}forecasting} models, a causal relationship between the predictors and the outcome may be required, while now-casting can also be achieved with predictive variables that do not necessarily have a causal effect on the outcome.
Several statistical models have been proposed for now-casting, for example hierarchical Bayesian models \citep{gunther2021nowcasting} or trend regression models \citep{kuchenhoff2021analysis}. Related approaches are discussed in \citet{altmejd2020nowcasting,schneble2021nowcasting,salas2021improving}.



\paragraph{Dynamic Models/Time-variant Dynamics}
A unique feature of pandemic assessment is the dynamic nature of the event. By this we not only refer to the explosive (exponential) growth that may occur but the fact that the properties of the processes that describe spatio-temporal changes are a function of time themselves. This is due to the fact that the behavior of the people continuously changes the properties of the system that we are trying to understand and make predictions for. This contrasts with other natural systems, like the current weather, and most systems in the engineering and physical sciences. 

Simple infectious disease compartmental models can be described by the stock of susceptible $S$, infected $I$, and removed population $R$ (either by death or recovery), 
\color{black} the contact rate $\kappa$, the infection probability $\beta$, the recovery rate $\gamma$, the death rate $\mu$ and the birth rate $\Lambda$ \citep{kermack1927contribution,hethcote2000mathematics,andersson2012stochastic,Grassly2008}. Here,

\begin{align*}
	\frac{dS}{dt} &= \Lambda - \mu S - \frac{\kappa \beta I S}{N}\\
	\frac{dI}{dt} &= \frac{\kappa \beta I S}{N} - \gamma I - \mu I\\
	\frac{dR}{dt} &= \gamma I - \mu R,
\end{align*}
where $t$ denotes the time point and $N$ is the total number of individuals in the population, i.e.~$N=I + S + R$.

Assuming that the susceptible individual first goes through a latent period after infection before becoming infectious $E$, adapted models such as SEI, SEIR or SEIRS, depending on whether the acquired immunity is permanent or not can be applied \citep{jit2011modelling}. Modeling the COVID-19 pandemic, applications include further approaches such as SIR-X accounting for the removal (quarantine) of symptomatic infected individuals and various other extensions and applications \citep{dehning2020inferring,lehr} {\color{black} including prediction of the impact of vaccination \citep{bubar2021}}.

In this deterministic compartment model, predictions are determined entirely by their initial conditions, the
set of underlying equations, and the input parameter values. Deterministic compartmental models have the advantage of being conceptually simple and easy to implement, but they lack for example stochasticity inherent in infectious disease transmission. In stochastic compartment models, the occurrence of events like transmission of infection or recovery is determined by probability distributions. Therefore, the chain of events (like an outbreak) is not exactly predictable. However, there are many possible types of stochastic epidemic models \citep{BRITTON201024, Kretzschmar2020}. 

{\color{black}\paragraph{Agent-based models (ABM)}
Agent-based modeling as an alternative approach uses individual-level simulation \citep{Karnon2012}. ABMs have been used to
model biological processes, ecological systems, traffic management, customer flow management or stock markets,
and in recent years increasingly for decision analysis as discussed later \citep{Marshall2015,Bonabeau2002,Macal2008}. ABMs
represent complex systems in which individual `agents' act autonomously and are capable of
interactions \citep{Miksch2019}. These agents can represent the heterogeneity of individuals, and the behavior
of individuals can be described by simple rules. Such rules include how agents interact, move
between geographical zones, form households or consume resources \citep{Chhatwal2015,Bruch2015,Hunter2017}. ABMs are often applied to study ``emergent behavior'' as a result of these predefined rules. In infectious disease modeling,
agent behaviors combined with transmission patterns and disease progression lead to
population-wide dynamics, such as disease outbreaks \citep{Macal2010}. In agent-based models, either all affected individuals are simulated individually, or specific networks of individuals are integrated into the simulation. }

{\color{black}\paragraph{Discrete Event Simulation (DES)}
Discrete event simulation is an individual-level simulation \citep{Pidd2004,Karnon2012,Jun1999,Zhang2018}. The core concepts of DES are entities (e.g.~patients), attributes (e.g.~patient characteristics), events, resources (i.e.~physical resources such as medical staff and medical equipment), queues and time \citep{Pidd2004,Banks2005,Jahn2010a}. In addition to health outcomes, performance measures such as resource use or waiting times can be calculated, as physical resources (e.g.~hospital beds) can be explicitly modeled \citep{Jahn2010a}. The term discrete refers to the fact that DES moves forward in time at discrete intervals (i.e.~the model jumps from the time of one event to the time of the next) and that events are discrete (mutually exclusive) \citep{Karnon2012}. 

\paragraph{Microsimulation}
Microsimulation methods, introduced by \citet{ Orcutt.1957}, are used to simulate policy actions on real populations.
\cite{Li.2013} describe microsimulations as ``a tool to generate synthetic micro-unit based data, which can then be used to answer many ``what-if''
questions that, otherwise, cannot be answered''.
The main difficulty for microsimulation is considered to be the choice of an appropriate data source on which these simulations can be conducted. Often, survey data are used. Nowadays, the first step in microsimulation is the realistic generation of data in the necessary geographic depth (e.g.\ \citealp{ Li.2013}). A full-population approach is described in \citet{mda2021.03}. Thereafter, the scenario-based microsimulation analysis yields the necessary information for building conclusions for policy support.
In microsimulation methods, we distinguish between static and dynamic models. The latter can be divided into time-continuous and time-discrete models. An overview of microsimulation methods is given in \citet{ Li.2013} and the references therein. For modeling COVID-19, dynamic models have to be considered. \citet{bock2020mitigation} presents a continuous time SIR microsimulation approach as an example for a dynamic transmission model. In contrast to ABM, other microsimulations are often based on survey data, or on realistic but synthetically extended survey data. The above-mentioned cohort simulations are usually deterministic simulations, in which an initial cohort of interest is followed over different paths over time, and thus leading to a distribution of outcomes after the analytic-time horizon. Recently, the dividing line between these methods and the related terminology has become blurred. Which method is ultimately used often depends on the background of the research team.}\\

{\color{black} A key} objective of forecasting in this context is to obtain numerically precise predictions, for variables such as the number of ICU beds. With this goal in mind, the reliability or accuracy of the predictions highly depends on the availability and quality of the data used to estimate the values of the parameters in the underlying mathematical or statistical models. 

It should be noted that these models are highly sensitive to context in the following sense: changes in the underlying system	in variables that are not part of the
	model can lead to changes in the relationship between the selected
	predictors and the predictions, rendering
	the predictions and their assumed
	uncertainty meaningless.





While it is widely appreciated that, for example, weather forecasts are only reliable for a couple of days, forecasting during a pandemic is even more complicated since the behavior of people influences the process that is being modeled. Forecasting during pandemics is, therefore, itself a continuous process with time-varying parameters. For this reason, such modeling effort is a complex undertaking requiring a range of data and expertise. Such activities should therefore be realized and coordinated through cross-disciplinary teams. To account for regional differences, one would expect a collective of modeling groups that support decision making for different parts of a country.

\hfill

\newpage

\subsection{Decision-analytic modeling}
Depending on the research question, different modeling approaches are used for decision-analytic modeling and development of computer simulations {\color{black} \citep{IQWIG2020, Roberts2012, STIKO2016}}. These include decision tree models, state-transition models, discrete event simulation models, agent-based models, and dynamic transmission models. \blue{Some of them have been introduced already since they are also commonly used for prediction. Models introduced in this section are predominantly used for decision analysis but could potentially be used for other purposes as well (Table \ref{tab:1}).
}

The selection of the model type depends on the decision problem and the disease. In general, decision trees are applied for simple problems, without time-dependent parameters and with a fixed and comparatively short time horizon. If the decision problem requires the evaluation over a longer time period and if parameters are time or age dependent, a state-transition cohort (Markov) models (STM) could be applied. STMs allow for the modeling of different health states and transitions between these states, and thus also for repeated events. They are applied when time to event is important. If the decision problem can be represented in an STM ``with a manageable number of health states that incorporate all characteristics relevant to the decision problem, including the relevant history, a cohort simulation should be chosen because of its transparency, efficiency, ease of debugging, and ability to conduct specific value of information analyses.'' \citep{Siebert2012}. If the representation of the decision problem  would lead to an unmanageable number of states, then an individual-level state-transition model is recommended \citep{Siebert2012}.  Especially in situations where interactions of individuals among each other or the health-care system need to be considered, that is, when we are confronted with scarce physical resources, queuing problems and waiting lines (e.g., limited testing capacities), discrete event simulation (DES) would be an appropriate modeling technique. DES allows the modeler to incorporate time-to-event data (e.g., time to progression) and physical resources are explicitly defined \citep{Karnon2012}. Modeling types such as differential equation systems, agent-based models and system dynamics account for the specific features of infectious diseases such as the transmissibility from infected to susceptible individuals and the uncertainties arising from complex natural history and epidemiology  \citep{Pitman2012,Grassly2008, jit2011modelling}.

\paragraph{Decision tree models}
In a decision-tree model, the consequences of alternative interventions or health technologies are described by possible paths. Decision trees start with decision nodes, followed by alternative choices (interventions, technologies, etc.) of the decision maker. For each alternative, the patients’ paths, which are determined by chance and that are outside the decision maker’s control, are then described by chance nodes. At the end of the paths, the respective consequences of each path are shown. Consequences or outcomes may include symptoms, survival, quality of life, number of deaths or costs. Finally, the expected outcomes of each alternative choice are calculated by taking a weighted average over all pathways \citep{Hunink2001,Rochau2015}, such as in the evaluation of COVID-19 testing strategies for university campuses in a decision-tree analysis \citep{Pelt2021}.

\paragraph{State-transition models}
A state-transition model is conceptualized in terms of a set of (health) states and transitions between these states. Time is represented in time intervals. Transition probabilities, time cycle length, state values (``rewards'') and termination criteria are defined in advance. During the simulations, individuals can only be in one state in each cycle. Paths of individuals determined by events during a cycle are modeled with a Markov cycle tree that uses a set of random nodes. The average number of cycles in which individuals are in each state can be used in conjunction with the rewards (e.g.~life years, health-related quality of life or costs) to estimate the consequences in terms of life expectancy, quality-adjusted life expectancy, and the expected costs of alternative interventions or health technologies. There are two common types of analyses of state-transition models: Cohort models (``Markov'') \citep{Beck1983,Sonnenberg1993} and individual-level models (``first order Monte Carlo'' models) \citep{Spielauer2007,GrootKoerkamp2010,Weinstein2006}.
Simple cohort models are defined in mathematical literature  as discrete-time Markov chains. A discrete-time Markov chain is a sequence of random variables ${\displaystyle X_{0},X_{1},X_{2},...}$ representing health states with the Markov property, namely that the probability of moving to the next health state depends only on the present state and not on the previous states:

${\displaystyle \Pr(X_{n+1}=x\mid X_{1}=x_{1},X_{2}=x_{2},\ldots ,X_{n}=x_{n})=\Pr(X_{n+1}=x\mid X_{n}=x_{n})}$

\begin{figure*}
	\includegraphics[width =  \textwidth]{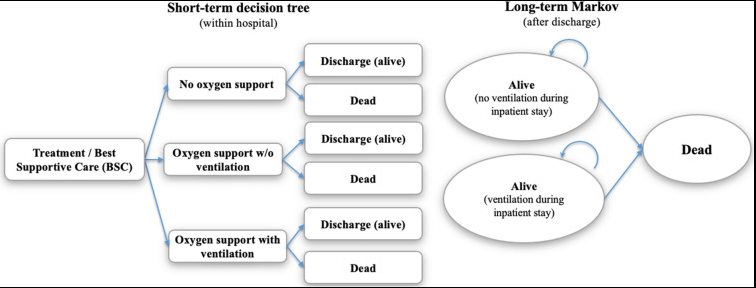}
	\caption{Example: A Cost-Effectiveness Framework for COVID-19 Treatments for Hospitalized Patients in the United States, \citep{Sheinson2021}}
	\label{fig:1}       
\end{figure*}

Generalized models such as continuous time Markov chains with finite or infinite state space are not commonly applied in health decision science. Applications of state-transition models in the pandemic include evaluation of treatments \citep{Sheinson2021} and vaccination strategies \citep{Kohli2021}. {\color {black} We also find hybrid models including the combination of decision trees and STMs (see Figure \ref{fig:1}).} 

\paragraph{Discrete Event Simulation (DES)}
 Similar to decision trees and state-transition models, health outcomes and costs of alternative health technologies can be assessed. {\color{black} In addition to these outcomes, as mentioned earlier, performance measures can be calculated as additional information for decision makers and the impact of scarce resources on costs and health outcomes can be evaluated \citep{Jahn2010}. The increased use of DES to support decision making under uncertainty is shown in the review of \cite{Zhang2018}.} Model applications in COVID-19 include optimizations of processes with scarce resources such as bed capacities \citep{melman2021balancing} or testing stations \citep{saidani2021designing} and laboratory processes \citep{gralla2020discrete}.

\paragraph{Dynamic Models/Time-variant Dynamics}
{\s The dynamic SIR type models explained in Section~\ref{sec:prediction} can also be used in the context of decision-analytic modeling {\color{black} \citep[e.g., vaccination allocation,][]{ecdc2020, Sandmann2021}} . This} model type can be extended by further compartments such as Death (D) and other states (X), reflecting, for example, quarantine or other states relevant to the research question. Such SIRDX models have been used frequently to model non-pharmaceutical intervention effects during the COVID-19 pandemics \citep{nussbaumer2020quarantine}. As in Markov state-transition models, a deterministic cohort simulation approach is used to model the distribution of compartments over time. Deterministic compartment models are useful for modeling the average behavior of disease epidemics in larger populations. When stochastic effects (e.g., the extinction of disease in smaller populations), more complex interactions between disease and individual behavior or distinctly nonrandom mixing patterns (e.g., the spread of the disease in different networks) are relevant, stochastic agent-based approaches can be used (see next section).

\paragraph{Agent-based models (ABM)}
{\color{black} Agent-based models as introduced earlier, have been used for decision analysis, for example for cost-effectiveness analyses in health care in the recent years \citep{Marshall2015, Chhatwal2015}. ABMs are also used in public health studies to model noncommunicable diseases \citep{nianogo2015agent}. }A comparison of ABM, DES and
system dynamics can be found in \cite{Marshall2015,Marshall2015a} and \cite{Pitman2012}. ABMs are increasingly applied for COVID-19 evaluations {\color{black}{including decision support for vaccination allocation accounting explicitly for network structure and contact behaviour \citep{Bicher2021, Jahn2021}.} 

{\color{black}\paragraph{Microsimulation}
Microsimulation as a modeling approach based on survey data and combining characteristics of above mentioned modeling approaches is used for prediction and decision analysis in various fields, especially for policy support using scenarios. Recently, MSM are also used for modelling diseases \citep{Hennessy2015} including infectious diseases.}}\\

Table \ref{tab:models} provides a short comparative overview of these commonly applied modeling approaches with example applications for COVID-19 research, in addition to our general comparison at the beginning of the section. Further guidance on model selection for a given problem at hand exists \citep{IQWIG2020,Roberts2012,Siebert2012, Marshall2015}. 


\begin{table}
	\caption{Overview of differences and similarities of simulation models commonly used as the basis for health decision sciences.}
	\label{tab:models}       
	\begin{tabularx}{\linewidth}{ X | X | X | X | X | X | X}
		Decision-Analytic Models & \makecell{Decision \\ trees} & State-transition models & Discrete event simulation & \makecell{Dynamic \\ models} & Agent-based models & \makecell{Micro-\\ simulation}\\ \hline
		\makecell{Mathe-\\matical /\\ statistical \\ background} & Operations research, decision analysis, machine learning & Markov chains, Monte Carlo methods & Monte Carlo methods & Differential equations (deterministic or stochastic) & Monte Carlo methods & Monte Carlo methods \\\hline
		Com\-ponents & Decision nodes, chance nodes, end nodes, paths & (Health) states, transition probabilities & Individuals (Entities), resources, event states & Compart\-ments, transition rates & Individuals (agents), attributes, learning rules, environment & Individuals, attributes, learning rules, environment \\ \hline
		Time & Not relevant & Time-dependent parameters & Discrete time intervals & Continuous & Continuous & Continuous\\ \hline
		Role and type of data & Model building/ Parameter estimation/ validation & Parameter estimation/ validation & Parameter estimation/ validation  & Parameter estimation/ validation & Parameter estimation/ validation & Survey data \\ \hline
		Other information & Expert opinion & Expert opinion & Expert opinion & Expert opinion & Expert opinion/ beliefs & \\ \hline
		Prominent application areas & Engineering, law, \black{health care} & Health care strategies & Queues & Infectious diseases & Biological processes & Traffic flow \\ \hline
		SARS-CoV-2 / COVID-19 examples & Testing strategies for university campuses \citep{Pelt2021} & Evaluation of vaccination strategies \citep{Kohli2021} & Balancing scarce resources \citep{melman2021balancing} & Contact tracing strategies \citep{Kretzschmar2020} & Vaccination strategies \citep{Jahn2021} & Herd immunity \citep{bock2020mitigation}\\
	\end{tabularx}
\end{table}

\section{Decision analysis}
\label{sec:decision}

The models described in {\s Section~\ref{sec:models1}} are built to inform decision making. Therefore, the so-called decision analysis framework is used.
Decision analysis aims to support decisions under uncertainty by means of systematic, explicit and quantitative methods. In particular, computer simulations and prediction models as described above are used to calculate the short-term and long-term benefits and harms (as well as the costs) of alternative interventions, technologies, or measures in health care \citep{Schoeffski2011, RSSDA2021}. The decision-analytic framework includes, among other things, the relevant health states and events considered to describe possible disease trajectories, the type of analysis (e.g., benefit-harm, cost-benefit, budget-impact analyses \citep{Drummond}), and the simulation method (cohort- or individual-based). In addition to base-case analysis (using the most likely parameters), scenario and sensitivity analyses \citep{Briggs2012} should be performed to show the robustness or uncertainty of the results. Value of information analysis can be applied to assess the value of future research to reduce uncertainty \citep{Fenwick2020,siebert2013enough}.

\subsection{Decision tradeoffs} \label{sec:tradeoff}
A central idea in decision analysis is that tradeoffs in outcomes of alternative choices are formalized and, if possible, quantified. In addition, the tradeoff between such outcomes is explicitly expressed, usually in the form of an incremental tradeoff ratio. In the context of a benefit-harm analysis, for example, this relates to quantifying the benefits of COVID-19 vaccination in terms of (incremental) deaths avoided and the harms of vaccination in terms of (incremental) potential side effects.{\color{black}Alternatively, the tradeoff of different school closure strategies in a pandemic (e.g., according to incidence level) weighting benefits in terms of (incremental) deaths or hospitalisations avoided and lost education time should be considered.} In general, two or more interventions can be compared in a stepwise incremental fashion \citep{keeney1976}. Benefit-harm analyses are often applied in screening evaluations {\color{black}\citep{Mandelblatt2016, Sroczynski2020}}. To detect efficient strategies, so-called strongly dominated strategies are first excluded. These are strategies that result in higher harms (e.g.~due to testing or invasive diagnostic work-up) and lower benefits (e.g., cancer-cases avoided, life-years gained) than other strategies. Second, weakly dominated strategies are excluded, that is strategies that result in higher harms per additional benefit compared with the next most harmful strategy, or in other words, strategies that are strongly dominated by a linear combination of any two other strategies. Third, the incremental harm-benefit ratios (IHBRs) are calculated for the non-dominated strategies. 

\begin{align*}
    IHB
    R &= \frac{\Delta harms}{\Delta benefits} &= \frac{harm (strategy_{i}) - harm (strategy_{i+1})}{benefit (strategy_{i}) - benfit (strategy_{i+1})} 
\end{align*}

There is no general benchmark for how much additional harm individuals are willing to accept per unit of additional benefit. Strategies are explored as a function of willingness-to-accept thresholds and they are displayed as harm-benefit acceptability curves on the efficiency frontier \citep{Neumann2016}. 
%
 
 In this context, the choice of measures that are presented and discussed also influences decision behavior \citep{Ariely2008}. The same applies to changes in decision-making due to alternatives that are presented. Regarding optimization of vaccination interventions, temporal aspects of availability and effectiveness of vaccinations can be considered. Alternative strategies can be evaluated, like in the comparison of immediate vaccination with lower vaccine protection, against later vaccination with expected higher effectiveness but risk of intermediate infection. {\color{black}Further non-pharmaceutical measures like school-closure strategies can be evaluated depending on incidence level but also accounting for additional measures to reduce the spread of the disease.}

\subsection{Statistical decision theory}

As a more general framework, statistical decision theory can help to make decisions on a formal basis. In this framework, the decision maker has to choose among a set of different actions $a$ by quantitatively assessing the consequences of these actions.
To this end, we consider a loss function $L(\theta, a)$, where the unknown parameter $\theta$ refers to the ``state of the world''.
The interesting question for the statistician is how to use the data in order to make optimal decisions. 
Assume we observe an experimental outcome $x$ with possible values in a set $\mathcal{X}$, which depends on the unknown parameter $\theta$. Furthermore, let $f(x | \theta)$ be the corresponding likelihood function.
Then, we define a decision function $\delta(x)$ which turns data into actions \citep{parmigiani2009decision}. To choose between decision functions,
we measure their performance by a risk function
$$
R(\theta, \delta) = \int_x L(\theta, \delta) f(x|\theta) dx.
$$
These can be approached from either a frequentist perspective (e.g.~the minimax decision rule) or a Bayesian perspective, where the risk is associated with a prior distribution $\pi(\theta)$. For a more thorough treatment of these concepts, we refer to \cite{parmigiani2009decision}.

Examples for loss functions in the context of COVID-19 include 
\blue{the number of avoided deaths \citep{bubar2021},}
negative reward functions in Markovian decision models \citep{eftekhari2020markovian} or
the social loss function as proposed in a recent discussion paper of the European Commission \citep{buelens2021lockdown}.

\section{Reporting and Communication}
\label{sec:reporting}


For data analysis and modeling to have an impact as a component of decision making, appropriate reporting and communication is key. There are {\color{black} numerous standards and guidelines for study planning and statistical reporting in the numerous application areas}, such as the ESS standard for quality reporting, the CONSORT, PRISMA, CHEERs guidelines, and others (see https://equator-network.org). These standards are based on commonly accepted core quality principles and values such as accuracy, relevance, timeliness, clarity, coherence, and reproducibility. For measures to restrain and overcome an epidemic effectively, communication among experts that follows highest professional and ethical standards is not sufficient. In a democratic society, policy measures can only be implemented if they are accepted by the wider population. This puts high demand on skills associated with communicating statistical evidence on the side of scientists, governments and media, and a citizenry able to understand statistical messages. 

In recent decades, there have been numerous publications, initiatives, and ideas to improve the communication of quantitative and statistical information, see \cite{Hoffrage2261,Tufte2001,Rosling2011,otavamylona2020}, to name only a few. Data journalism has recently taken off as an innovative component of news publishing, and COVID-19 provides numerous excellent examples, often using an interactive visual format on the Internet, such as dashboards. 
A fundamental problem in assessing probabilities, for example, lies in the intuitive conflation of subjective risks (``how likely am I to become infected'') and general risks (``how likely is it that some person will become infected''). Another issue is that of equating sensitivity of a diagnostic test and the positive predictive value \citep{Eddy1982,Gigerenzer2007,McDowell2019,Binder2020}. In particular, the prevalence (or base rate) is often neglected leading to this confusion. 
Fact boxes combined with icon arrays are recommended for the presentation of test results. Both representations are based on natural frequencies \citep{[62],Krauss2020} and present case numbers as simply and concretely as possible. Many scientific studies show that icon arrays help people understand numbers and risks more easily \citep[e.g.][]{McDowell2019}. The Harding Center for Risk Literacy shows many other examples of transparent communication of risks, including COVID-19\footnote{\url{https://www.hardingcenter.de/de/}} and {\color{black}a collection of misleading or wrong communication of statistics, such as for vaccination effects ("Unstatistik'')}.

\begin{figure*}
	\includegraphics[width = 0.9\textwidth]{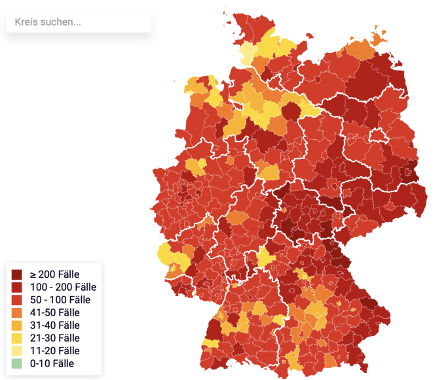}
	\caption{Choropleth map of the incidence figures \blue{on 2021-02-12} for Germany by district. Source: Robert-Koch-Institute https://app.23degrees.io/export/oCRP768wQ3mCswE7-choro-corona-faelle-pro-100-000/image.}
	\label{fig:choro}
\end{figure*}

{\color{black} When communicating the results of an analysis to policy makers or the general public, the following aspects must also be kept in mind.} While human thinking tends towards pattern simplification and political communication also prefers a simple cause-effect relationship, real phenomena are often multivariate. Thus, when studying COVID-19 and predicting its spread, it is important to consider its symptomatology, the incidence and geographic distribution of diseases, population behavior patterns, government policies and impacts on the economy, on schools, on people in nursing homes and on social life as a whole. However, it is also crucial to integrate these into data analyses and to communicate results clearly and transparently; for example, it might be important to state that associations observed in the data could be caused by other, omitted variables (confounders). In addition, much of the data comes from observational studies, which usually makes a robust causal attribution problematic. 
{\s As many of these phenomena cannot be studied other than by observation (for ethical and feasibility reasons), causal attribution might be achieved as a scientific consensus opinion among scientists from the relevant disciplines that understand the complexity of the models and the subject matter studied.
}
Visual representations take a central position in public communication and aim to represent the corresponding dynamics and contents in a quickly understandable way. Usually either time-dependent parameters or data with a spatial reference are visualized. For spatially distributed data, choropleth maps are predominantly used, in which administrative regions defined by the responsible health authorities are colored according to the distribution density of the infection figures or variables derived from them (see Figure~\ref{fig:choro}).  Their visual perception problems - such as the visual dominance of the area of administrative regulatory frameworks that have no direct relation to infection events - are well known but the effects of such problems are still widespread.  In addition, the use of ordinance thresholds as the basis for color scaling is often at odds with color schemes that emphasize real spatial distributional differences.

\begin{figure}
\centering
		\includegraphics[width = 0.7 \textwidth]{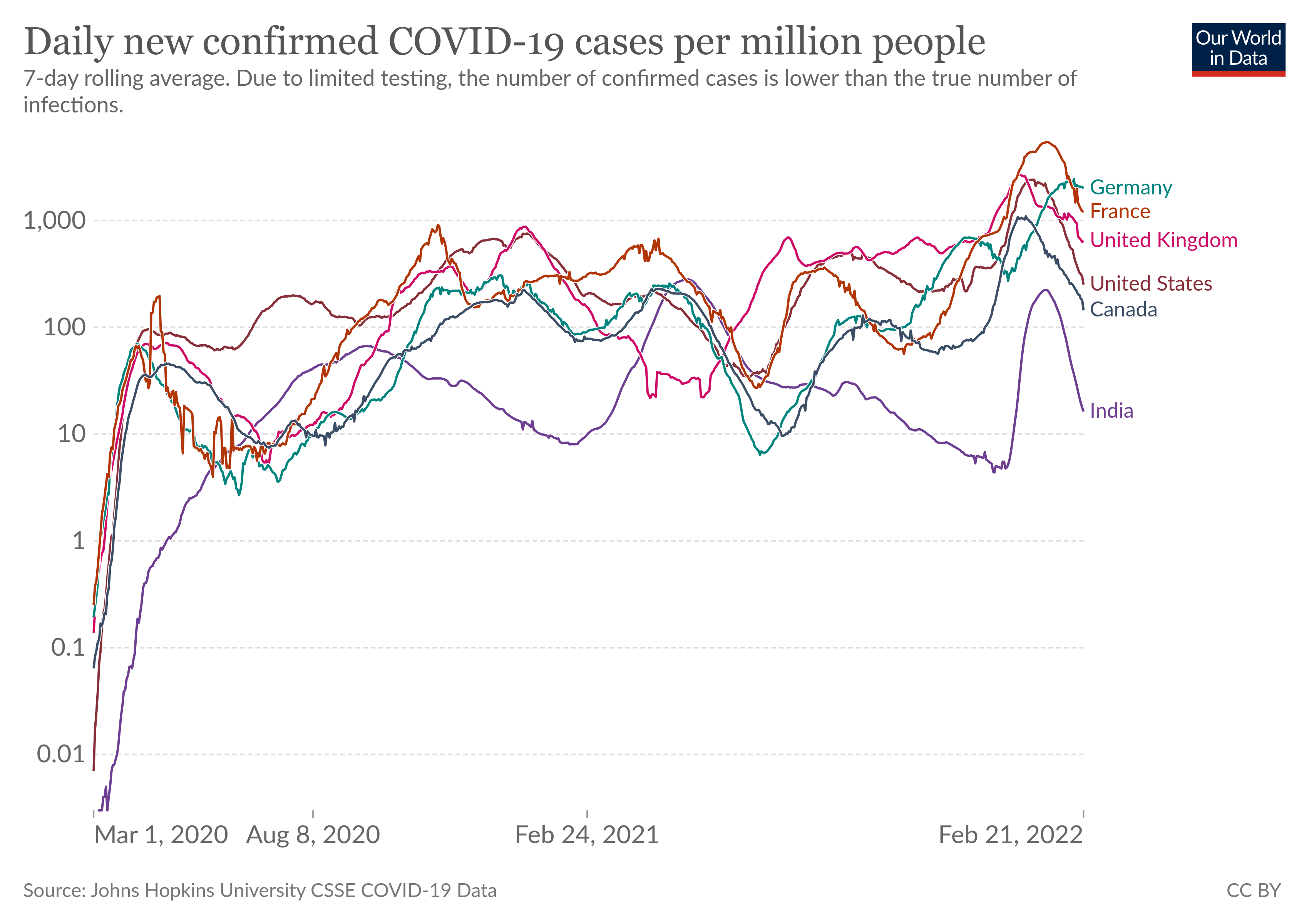}
			\includegraphics[width = 0.7 \textwidth]{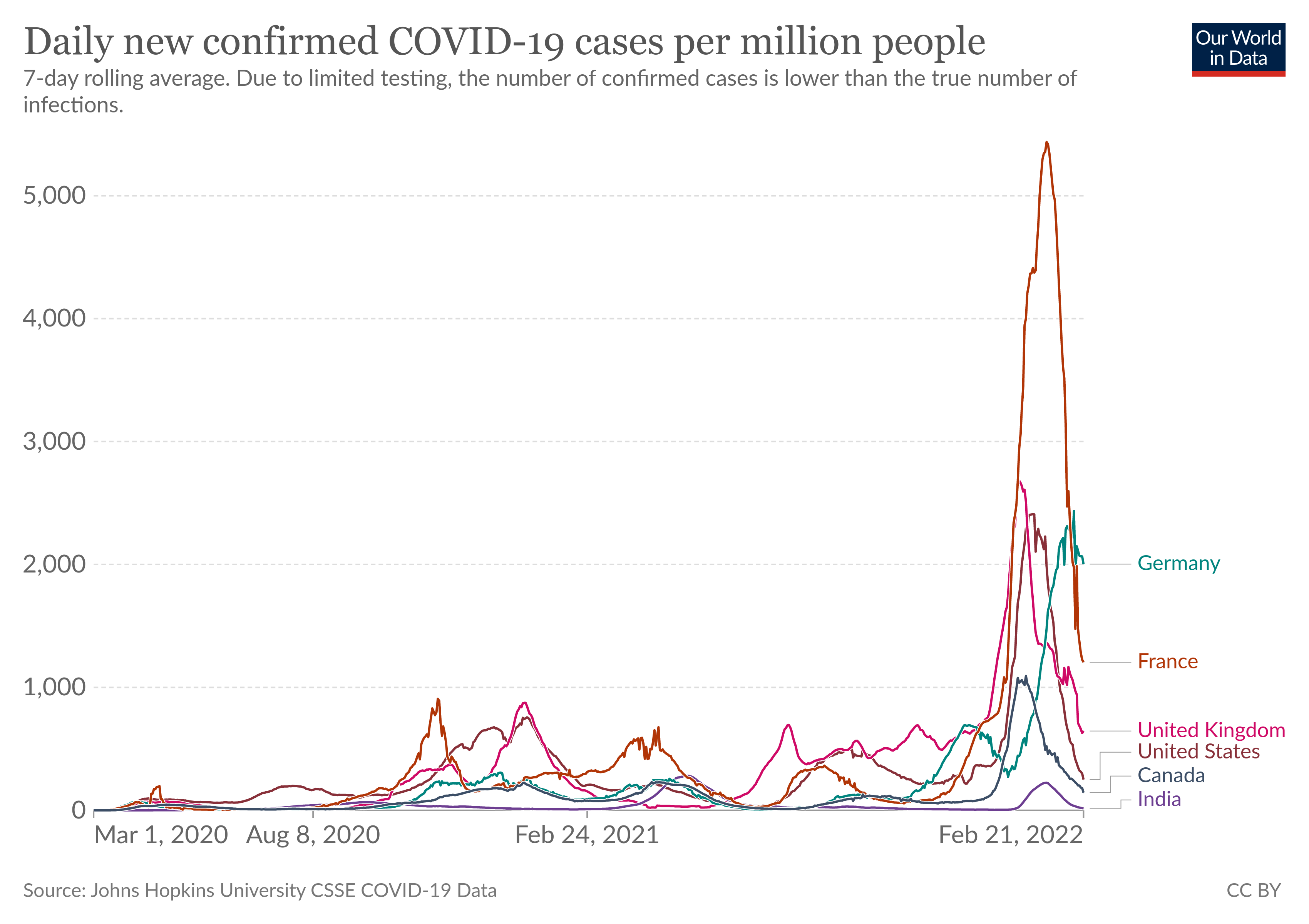}
	\caption{The 7-day incidence for different countries over time. On the logarithmic scale ({\color{black} top} graphic), differences appear small. The linear scale ({\color{black}bottom} graphic), however, shows considerable differences. Source: Our World in Data, https://ourworldindata.org/covid-cases?country=IND~USA~GBR~CAN~DEU~FRA, \blue{accessed 2022-02-22}.}
	\label{fig:incidence}
\end{figure}

For time-dependent parameters, different variants of time series diagrams are used, predominantly line and column diagrams. 
The use of logarithmic scales in time series diagrams should be evaluated with caution \citep{romano2020}. On the one hand, they tempt superficial readers to underestimate dynamic growth processes; on the other hand, they increase the demands on the mathematical and statistical literacy of the readership without corresponding advantages of visual representation. Figure~\ref{fig:incidence} shows the time course of the 7-day incidence per 100,000 people between 24 January and 4 February 2021 for some selected countries. While the differences appear relatively small on the logarithmic scale, the linear scale shows considerable differences.

{\color{black} 
\section{Recommendations} \label{sec:recom}
}
Reaching a decision based on data requires several steps, which we have illustrated in this paper: Data provide the basis for different kinds of models, which can be used for prediction, explanation and decision making. This forms the basis for making decisions within a formal framework. The results of these models must be communicated to non-scientists in order to gain acceptance of and adherence to policy decisions. Each of these steps comes with its own caveats and requires sound statistical knowledge. 

{\color{black}
\textbf{Data:} Lessons learned from the current pandemic about data, variables and information that should be obtained are critically discussed for specific countries \citep{Grossmann2022, RSS, rendtel2021} and on a European and international level \citep{KUCHARSKI2021, ecdc2020Performance}, {\color{black}\citep{dean,Mathieu2022}}. Examples include establishing new vaccination registries, extending coronavirus registries with further socio-demographic parameters and data sharing. \textbf{We recommend} the implementation of standards \citep{european} and processes for data collection on a national and international level, especially within Europe. These standards need to be refined, meeting the requirements of relevance, transparency, truthfulness, timeliness and accuracy to improve the handling of epidemics and pandemics in the coming years. National and international strategies and systematic collection and sharing of data allowing researchers to access the information is important to build comparable statistical and decision-analytic models. 

\textbf{Modeling Methods:} Depending on the modeling approach, the model can be fit to one data set (e.g., regression model) or data from different sources with different levels of evidence can be used to populate the model (e.g., decision-analytic models). In addition, several modeling approaches may be applicable for one research question (e.g., differential equation model or ABM for the prognosis of COVID-19 spread and consequences). \textbf{We recommend} clear communication about 1) the purpose of the model, 2) how the model uses data, 3) the database or additional assumptions and their evidence basis, and 4) risks and uncertainties. If applicable, several modeling approaches should be applied. As a result, the model best fitting the data would be selected or different modeling approaches could provide insights into uncertainty, like in national forecasting consortia (e.g., as in the Austrian COVID Prognosis Consortium \citep{ProgAustria}) and nowcast/forecast ensembles. The infrastructure of comprehensive population or microsimulation models – including population, disease and flexible intervention or policy modules – needs to be established and maintained beyond the current crisis.

\textbf{Data Aquisition:} \color{black}
Models require data from various sources and different modeling approaches allow data transformation and synthesis from different sources. Data aquisition for scientific evaluations requires a further improved infrastructure to speed up model development and to parameterize models with high level evidence {\color{black} including vaccine effectiveness in real time}.  \textbf{We recommend} the creation of a central national DataLab, \blue{collecting data in a unified way and linking} data from different sources \blue{as well as} enabling accessibility for experienced users. With regard to data sharing, it is important that this is manageable from a practical point of view in terms of the time frame and resources needed.

\textbf{Transparency:} In the wake of the COVID-19 pandemic, reporting of infection numbers and derived epidemiological indicators boomed, demonstrating with dramatic clarity the knowledge gap between experts, policymakers, and the public. 
 To increase the acceptance of decisions and associated measures, all steps in the decision-making process must be disclosed. \textbf{We recommend} a transparent decision-making process and communication of this process starting with the data, continuing with the choice of models, relevant perspectives, outcomes or metrics for several outcomes and an explicit discussion of considered tradeoffs \citep[see, e.g.,][]{Gigerenzer2007,gigerenzer2003simple,RSSDA2021}. In this context, the media also play a crucial role.

\textbf{Interdisciplinary cooperation:} A pandemic poses particular challenges to society as a whole. In order to tackle these as efficiently as possible, interdisciplinary cooperation, such as that fostered by the DAGStat, is essential. 
\textbf{We recommend} that experts act as a specialist group rather than as individuals, broadly positioned and media-sensitive. These interdisciplinary collaborations should consist of data scientists including statisticians, epidemiologists, experts in public health, social sciences, and ethics, as well as decision and communication scientists. The DAGStat as an umbrella organization of various professional societies, the Competence Network Public Health COVID-19 or the Society for Medical Decision Making (SMDM) are examples for existing networks that can be built upon. 

\textbf{Statistical Literacy / Data Literacy:} The COVID-19 crisis brought into the general public's awareness that our social interaction and political decisions are essentially based on data, modeling, the weighing of risks and benefits, and thus on probability estimates, expected values and incremental harm-benefit ratios. Clearly, we need additional efforts to promote statistical or data literacy at all levels of society. 
 The ability to critically evaluate and interpret data and to critically reflect on model outcomes serves to promote maturity in a modern digitized world. 
\textbf{We recommend} promotion of statistical literacy to be intensified at all levels of education (school/vocational education/training) following the Data Literacy Charta \citep{DataLiteracyCharta} and including risk competency \citep{CAPUR2019, Loss}.  Therefore, collaboration between statisticians and all stakeholders involved in statistical literacy is necessary.\\
}

\section{\color{black} Discussion}
\label{sec:discussion}
	

{\color{black}
In our paper, we discussed all steps starting from data capturing to statistics, modeling, decision making and communication which are important aspects in the context of evidence-based decision making. The current pandemic has shown that, in particular in Germany, we are still far from such an evidence-based decision-making process. 
Aims of this process include the following: First, it should result in the best possible decision given the available evidence and it is necessary to explicitly consider the tradeoffs involved with certain interventions. Second, gaining the public's acceptance of the decisions is fundamental.
In order to achieve these goals, we need reliable data, careful interpretation of the results and a clear communication, especially concerning uncertainty, see, for instance, \cite{who}.}

It is important to note that the considerations described in our paper not only apply to the current pandemic, but also extend to future pandemics\footnote{\url{https://www.statnews.com/2021/05/18/luck-is-not-a-strategy-the-world-needs-to-start-preparing-now-for-the-next-pandemic/}} and other (public or political) challenges such as the climate change debate \blue{\citep{Ritchie2021}}.
	
Our paper has several limitations. First, we refer to data as the input to the statistical models irrespective of possible preprocessing steps. An upcoming publication on data and data infrastructure in Germany will include more details on data preparation, building upon the DAGStat white paper \citep{DAGStat2021}. Second, our overview of modeling techniques does not provide a detailed discussion of all modeling approaches and their advantages and limitations but it should foster interdisciplinary collaboration among data scientists. References on guidance papers provide valuable further readings. 
{\color{black}Third, the decision-making process involves a variety of stakeholders, including politicians, government agencies and health authorities, health care providers, citizens, patients and their relatives, scientists, and they all take different perspectives. We have not discussed this aspect in detail in our paper, but the original white paper included a paragraph on political decision making \citep{DAGStat2021}.
}

The German Consortium in Statistics (DAGStat), a network of 13 statistical associations and the German Federal Office of Statistics\footnote{\s \url{https://www.dagstat.de/en/about-us/cooperating-societies}} and {\color{black}the Society for Medical Decision Making (SMDM)}, initiated a collaboration of scientists with backgrounds in all areas of statistics as well as epidemiology, decision analysis and political sciences to critically discuss the role of data and statistics as a basis for decision-making motivated by the COVID-19 pandemic. We found that similar concepts are often considered in different areas, but different notation and wording can hinder transferability. In this sense, this paper also aims to bridge the gaps between disciplines, and to broaden the research focus of statistical disciplines to prepare for future pandemics.

\begin{acknowledgements}
We thank Lyndon James (Harvard T.H. Chan School of Public Health) for proofreading and language editing.
\end{acknowledgements}

%
\section*{Funding}
Tim Friede and Markus Pauly are grateful for support by the Volkswagen Foundation (``Bayesian and Nonparametric Statistics - Teaming up two opposing theories for the benefit of prognostic studies in Covid-19''). Research by Beate Jahn and Uwe Siebert was also funded in part by the Austrian Federal Ministry for Digital and Economic Affairs BMDW and handled by the Austrian Research Promotion Agency (FFG) within the Emergency Call for research into COVID-19 in response to the SARS-CoV-2 outbreak (CIDS—Concurrent Infectious Disease Simulation) (881665) and by the Gordon and Betty Moore Foundation through Grant (GBMF9634) to Johns Hopkins University to support the work of the Society for Medical Decision Making COVID-19 Decision Modeling Initiative. 

 \section*{Conflict of interest}

 The authors declare that they have no conflict of interest.

\section*{Availability of data and material}

Not applicable.

\section*{Code availability}
Not applicable.

\section*{Authors' contributions}

\bibliographystyle{spbasic}      
\bibliography{Lit-DAGStat-COvid.bib}   

\end{document}